\PassOptionsToPackage{pdftex}{graphicx}
\documentclass[letterpaper,10pt,authoryear,times,twocolumn,5p,lefttitle,final]{elsarticle} 
\usepackage[utf8]{inputenc}
\usepackage[T1]{fontenc}
\usepackage[pdfencoding=auto,hidelinks]{hyperref}
\usepackage[dvipsnames]{xcolor}
\usepackage{amsmath}
\usepackage{array}
\usepackage{booktabs}
\usepackage{colortbl}
\usepackage{siunitx}
\usepackage{multirow}
\usepackage{textgreek}
\usepackage{textcomp}
\usepackage{subcaption}
\usepackage{scalerel}
\usepackage{tikz}
\usepackage{gensymb}
\usepackage{pgfplots}
\pgfplotsset{compat=1.16}
\usepgfplotslibrary{fillbetween,groupplots}
\usepackage{pgfplotstable}
\usepackage{xfrac}
\DeclareInstance{xfrac}{cmr}{text}{slash-symbol-font = ptm}
\usepackage{svg}
\DeclareRobustCommand{\orcid}[1]{\texorpdfstring{\raisebox{3pt}{\href{https://orcid.org/#1}{\includesvg[width=5pt]{figures/assets/orcid}}}}{}}

\newcommand{\chiv}{CH\textsubscript{4}}
\newcommand{\coii}{CO\textsubscript{2}}
\newcommand{\ppmm}{ppm$\cdot$m}
\renewcommand{\vec}[1]{\boldsymbol{#1}}

\makeatletter
\def\ps@pprintTitle{%
     \let\@oddhead\@empty
     \let\@evenhead\@empty
     \def\@oddfoot
       {\hbox to \textwidth%
        {\ifnopreprintline\relax\else
        \@myfooterfont%
         \ifx\@elsarticlemyfooteralign\@elsarticlemyfooteraligncenter%
           \hfil\@elsarticlemyfooter\hfil%
         \else%
         \ifx\@elsarticlemyfooteralign\@elsarticlemyfooteralignleft%
           \@elsarticlemyfooter\hfill{}%
         \else%
         \ifx\@elsarticlemyfooteralign\@elsarticlemyfooteralignright%
           {}\hfill\@elsarticlemyfooter%
         \else%
               \ifx\@journal\@empty%
                 Elsevier%
            \else\@journal\fi\hfill\fi%
         \fi%
         \fi%
         \fi%
         }
       }%
     \let\@evenfoot\@oddfoot}
\makeatother
\journal{{\scriptsize Accepted Manuscript for Remote Sensing of Environment --- Special Issue: Remote Sensing of Greenhouse Gas Emissions --- \href{https://doi.org/10.1016/j.rse.2021.112574}{doi: 10.1016/j.rse.2021.112574} --- \href{http://creativecommons.org/licenses/by-nc-nd/4.0/}{CC BY-NC-ND}}}

\makeatletter
\gdef\emailauthor#1#2{\stepcounter{ead}%
     \g@addto@macro\@elseads{\raggedright%
      \let\corref\@gobble\def\@@tmp{#1}%
      \eadsep{\ttfamily\expandafter\strip@prefix\meaning\@@tmp}
      \def\eadsep{\unskip,\space}}%
}
\makeatother

\begin{document}
\begin{frontmatter}
\title{Impact of Scene-Specific Enhancement Spectra on Matched Filter Greenhouse Gas Retrievals from Imaging Spectroscopy}
\author[1,2]{Markus D. Foote\orcid{0000-0002-5170-1937}}
\author[3]{Philip E. Dennison\orcid{0000-0002-0241-1917}}
\author[3]{Patrick R. Sullivan\orcid{0000-0003-3656-2518}}
\author[3]{Kelly B. O'Neill\orcid{0000-0001-8935-8057}}
\author[4]{Andrew K. Thorpe\orcid{0000-0001-7968-5433}}
\author[4]{David R. Thompson\orcid{0000-0003-1100-7550}}
\author[4]{Daniel H. Cusworth\orcid{0000-0003-0158-977X}}
\author[4,5]{Riley Duren\orcid{0000-0003-4723-5280}}
\author[1,2]{Sarang C. Joshi}
\address[1]{Scientific Computing and Imaging Institute, University of Utah, Salt Lake City, UT, USA}
\address[2]{Department of Biomedical Engineering, University of Utah, Salt Lake City, UT, USA}
\address[3]{Department of Geography, University of Utah, Salt Lake City, UT, USA}
\address[4]{Jet Propulsion Laboratory, California Institute of Technology, Pasadena, CA, USA}
\address[5]{Office of Research, Innovation, and Impact, University of Arizona, Tucson, AZ, USA}

\begin{abstract}
Matched filter techniques have been widely used for retrieval of greenhouse gas enhancements from imaging spectroscopy datasets.
While multiple algorithmic techniques and refinements have been proposed, the greenhouse gas target spectrum used for concentration enhancement estimation has remained largely unaltered since the introduction of quantitative matched filter retrievals. 
The magnitude of retrieved methane and carbon dioxide enhancements, and thereby integrated mass enhancements (IME) and estimated flux of point-source emitters, is heavily dependent on this target spectrum. 
Current standard use of molecular absorption coefficients to create unit enhancement target spectra does not account for absorption by background concentrations of greenhouse gases, solar and sensor geometry, or atmospheric water vapor absorption. 
We introduce geometric and atmospheric parameters into the generation of scene-specific unit enhancement spectra to provide target spectra that are compatible with all greenhouse gas retrieval matched filter techniques. 
Specifically, we use radiative transfer modeling to model four parameters that are expected to change between scenes: solar zenith angle, column water vapor, ground elevation, and sensor altitude.
These parameter values are well defined, with low variation within a single scene. 
A benchmark dataset consisting of ten AVIRIS-NG airborne imaging spectrometer scenes was used to compare IME retrieved using a matched filter algorithm. 
For methane plumes, IME resulting from use of standard, generic enhancement spectra varied from $-$22~to~$+$28.7\% compared to scene-specific enhancement spectra. Due to differences in spectral shape between the generic and scene-specific enhancement spectra, differences in methane plume IME were linked to surface spectral characteristics in addition to geometric and atmospheric parameters. 
IME differences were much larger for carbon dioxide plumes, with generic enhancement spectra producing integrated mass enhancements $-$76.1~to~$-$48.1\% compared to scene-specific enhancement spectra. 
Fluxes calculated from these integrated enhancements would vary by the same percentages, assuming equivalent wind conditions. Methane and carbon dioxide IME were most sensitive to changes in solar zenith angle and ground elevation.
We introduce an interpolation approach that can efficiently generate scene-specific unit enhancement spectra for given sets of parameters. 
Scene-specific target spectra can improve confidence in greenhouse gas retrievals and flux estimates across collections of scenes with diverse geometric and atmospheric conditions.

\newline\\[-2mm]
\begin{minipage}{\textwidth}
\textit{Research Highlights:} 
\vspace{-1.5mm}
\begin{enumerate}
    \item Matched filter greenhouse gas retrievals should account for atmosphere and geometry
    \item A benchmark imaging spectroscopy dataset facilitates comparison of retrieval methods 
    \item Standard approach has $-$22 to +29\% of \chiv{} mass compared to scene-specific approach 
    \item Standard approach omits 48.1--76.1\% of \coii{} mass compared to scene-specific approach
    \item \chiv{} and \coii{} retrievals are most sensitive to solar zenith angle and ground elevation
\end{enumerate}
\end{minipage}
\vspace{-2.25mm}
\end{abstract}

\end{frontmatter}
\newpage
\section{Introduction}
Imaging spectrometer data with 5--10~nm spectral resolution have become widely used for greenhouse gas point-source detection and emissions quantification \citep{Dennison2013,Thompson2015,Frankenberg2016,Thompson2016,Thorpe2017,Duren2019,Cusworth2021}. Two types of algorithms have commonly been used to retrieve point source methane (\chiv{}) and carbon dioxide (\coii{}) enhancements from imaging spectrometer data: matched filters and differential optical absorption spectroscopy (DOAS). DOAS retrievals are computationally intensive, and have focused on estimating \chiv{} and \coii{} concentrations in small areas surrounding known point-source emitters \citep{Thorpe2014,Frankenberg2016,Thorpe2017,Cusworth2019}. Matched filters, in contrast, are applied to retrieve greenhouse gas enhancements over entire scenes and flight campaigns, regularly surpassing multiple terabytes of data, and used as a screening tool for tactical operations within a campaign \citep{Thompson2015,Frankenberg2016,Thompson2016,Ayasse2019,Duren2019,Foote2020,Thorpe2020}. Matched filter data products have been utilized in subsequent analyses of \chiv{} source distributions, including flux estimation and source attribution \citep{Frankenberg2016,Duren2019}. Refinements to matched filter methods have largely focused on improving computational efficiency and removing instrument effects \citep{Manolakis2014a,Thompson2015,Foote2020}. However, the accuracy of matched filter retrieval algorithms depends on the target spectrum that is used as a `search query' to identify the signature of enhanced trace gas absorption in imaging spectrometer data.

For matched filter analysis, imaging spectrometer data are conventionally interpreted as a data cube in which two dimensions represent the spatial extent of the data, and the third dimension encodes the sensor-measured reflected solar radiance at varying wavelengths, \textit{i.e.}, a radiance spectrum.
In its simplest form, the matched filter operates only on this spectral dimension, the values of which we denote in a data vector $\vec{x}$. 
For an individual radiance spectrum, the linearized matched filter takes the form
\begin{equation}
    \alpha = \frac{(\vec{x}-\vec{\mu})^T \vec{\Sigma}^{-1} \vec{t}}{\vec{t}^T\vec{\Sigma}^{-1}\vec{t}}
    \label{eq:mf}
\end{equation}
where $\vec{\mu}$ is the mean radiance, and $\vec{\Sigma}$ is the covariance of radiance within the dataset that whitens the data \citep{Theiler2006a,Theiler2006b,Manolakis2009b,Theiler2012}.
The target spectrum $\vec{t}$ determines the spectral features to which the matched filter is sensitive.
For trace gas detection, the target spectrum is constructed to represent the characteristic absorption features for a molecular species. In quantitative retrievals, a \textit{unit enhancement spectrum} represents a change in transmittance corresponding to a change in concentration, expressed as parts per million (ppm) of enhanced concentration over a pathlength in meters (m) \citep{Thompson2015}. The resulting unit enhancement spectrum has units of (\ppmm)\textsuperscript{-1}. When multiplied by the mean radiance within an area, the resulting target spectrum has units of radiance per \ppmm{} (\textit{e.g.}, \textmu{}W\,cm\textsuperscript{-2}\,sr\textsuperscript{-1}\,nm\textsuperscript{-1}\,(\ppmm)\textsuperscript{-1}). Thus, attenuation is proportional to the background radiance, and changes the shape and magnitude of the target spectrum. Quantitative matched filter retrievals output an image containing trace gas concentration-path length enhancement, in \ppmm.  

Previous applications of quantitative matched filter retrievals have primarily used a standard, generic unit enhancement spectrum for \chiv{} \citep{Thompson2015,Frankenberg2016,Thompson2016,Krautwurst2017,Duren2019,Thorpe2020}. This approach uses the absorption coefficients from the HITRAN molecular absorption database \citep{Rothman2013} and a simple forward radiative transfer simulation based on a 1 m layer of \chiv{} at standard temperature and pressure \citep{Thompson2015}. Generic unit enhancement spectra based on this method will not account for absorption by the background concentration of trace gases, including ambient \chiv{}, \coii{}, and water vapor. Both \chiv{} and \coii{} absorption features in the shortwave infrared partially overlap with water vapor absorption. Background concentration is especially concerning for potential creation of a \coii{} unit enhancement spectrum, since \coii{} has a higher background concentration relative to \chiv{}. The generic unit enhancement spectrum method also assumes one-directional transmittance along a path, and does not correct for two-directional transmittance (e.g. downwelling and upwelling) or adjust for solar zenith angle. Considering a plume as layer of enhanced concentration, higher solar zenith angles will greatly increase pathlength, such that a point source plume in a high solar zenith angle imaging spectrometer scene will return higher retrieved enhancements (and result in a higher estimated flux rate) than the same point source plume in a low solar zenith angle scene, if a generic unit enhancement spectrum is used. 

Calculating scene-specific unit enhancement spectra can account for background concentration of trace gases and geometric effects, thereby increasing the accuracy of both enhancement retrievals and flux estimation. In this paper, we introduce a simulation strategy to determine the scene-specific unit enhancement spectra of \chiv{} and \coii{}, based on five key parameters: background concentration, solar zenith angle, sensor altitude, ground elevation, and column water vapor. We then analyze the importance of scene-specific unit enhancement spectra through a comparison using a diverse benchmark dataset consisting of ten imaging spectrometer scenes containing \chiv{} and \coii{} plumes. We present a method for portable generation of approximate unit enhancement spectra using a lookup table and interpolation, followed by determination of the error in this approximation. Finally, a sensitivity analysis determines the sensitivity of retrievals to four geometric and atmospheric parameters.

\section{Methods}
This section describes our approach for generating scene-specific unit enhancement spectra and analysis of the effects of this approach versus the previous generic unit enhancement spectrum approach. In section~\ref{sec:methods:dataset} we describe the selection of scenes used for analysis and the tabulation of associated metadata for the selected geometric and atmospheric parameters. Sections~\ref{sec:methods:generic} and \ref{sec:methods:simulation} then describe unit enhancement spectrum generation techniques from a generic library and by using our novel approach with scene-specific geometric and atmospheric parameters.
The scene-specific method is made more portable with the interpolation approach described in section~\ref{sec:methods:lut}.
Section~\ref{sec:methods:ime} details the calculation of integrated mass enhancement used for comparison of unit enhancement spectrum techniques for \chiv{} and \coii{} enhancement retrievals.
Finally, section~\ref{sec:methods:sens} outlines an experiment for approximating the sensitivity of IME calculations to each geometric/atmospheric parameter.


\newcommand*\rot{\rotatebox{90}}
\newcommand{\sceneloc}[1]{\begin{minipage}[t]{3.4cm}\raggedright #1\end{minipage}}

\setlength{\tabcolsep}{3pt}
\renewcommand{\arraystretch}{1.5}
\newcolumntype{L}[1]{>{\raggedright\let\newline\\\arraybackslash\hspace{0pt}}p{#1}}
\begin{table*}[t]
    \centering
    \caption{Benchmark Dataset Scenes Plume Summary and Geometric/Atmospheric Parameters}
    \label{tab:scene_table}\footnotesize
    \begin{tabular}{
        l
        >{\raggedright}p{3.25cm}
        S[table-format=2.1,round-mode=figures,round-precision=3]
        S[table-format=1.2,round-mode=places,round-precision=2]
        S[table-format=1.2,round-mode=places,round-precision=2]
        S[table-format=1.2,round-mode=places,round-precision=2]
        S[table-format=1.1,round-mode=places,round-precision=1]
        L{2.92cm}
    }
        \toprule
        {\footnotesize\textrm{\shortstack[l]{Scene Identifier\\and Location}}}&
        \shortstack[c]{Plume Features} &
        \shortstack[c]{Zenith\\(degrees)} &
        \shortstack[c]{Water\\(cm)} &
        \shortstack[c]{Ground\\(km)} &
        \shortstack[c]{Sensor\\(km)} &
        \shortstack[c]{Resolution\\(m)} &
        \shortstack[c]{Prior Investigation} \\
        \midrule
        \sceneloc{\texttt{ang20150420t181345}\\New Mexico, USA           } & Coal mine vent \chiv{}, coal power plant \coii{}     & 28.387218 & 1.23666  & 1.632398687 & 2.816333451 & 1.1 & \citet{Frankenberg2016,Thorpe2017}\\
        \sceneloc{\texttt{ang20160211t075004}\\Ahmedabad, Gujarat, India } & Landfill, wastewater treatment, and oil pump \chiv{} & 37.71064  & 1.76964  & 0.044983918 & 8.455075394 & 8.1 & \citet{Foote2020} \\ 
        \sceneloc{\texttt{ang20170616t212046}\\Corcoran, California, USA } & Dairy waste digester \chiv{}                         & 21.786715 & 4.30692  & 0.066331916 & 3.170339244 & 3.0 & \citet{Ayasse2019,Duren2019}\\
        \sceneloc{\texttt{ang20170618t193955}\\San Jose, California, USA } & Landfill and wastewater treatment \chiv{}            & 15.332411 & 5.47161  & 0.011119466 & 3.443644896 & 3.3 & \citet{Ayasse2019,Duren2019}\\
        \sceneloc{\texttt{ang20170906t210217}\\Elk Hills, California, USA} & Oil and natural gas extraction \chiv{}               & 32.916242 & 4.79577  & 0.353131316 & 2.484402521 & 2.1 & \citet{Duren2019}\\
        \sceneloc{\texttt{ang20180415t045439}\\Maithon, Jharkhand, India } & Coal power plant \coii{}                             & 23.420029 & 3.5221   & 0.155032913 & 4.367060251 & 4.0 & \citet{Cusworth2021}\\
        \sceneloc{\texttt{ang20180927t195028}\\San Jose, California, USA } & Landfill and wastewater treatment \chiv{}            & 39.34137  & 3.08051  & 0.00582665  & 3.370477798 & 3.2 & --\\
        \sceneloc{\texttt{ang20190621t200919}\\Fruitland, New Mexico, USA} & Coal power plant \coii{}                             & 17.558022 & 1.59725  & 1.609315396 & 4.429259552 & 2.7 & \citet{Cusworth2021}\\
        \sceneloc{\texttt{ang20191004t221515}\\Thompsons, Texas, USA     } & Coal power plant \coii{}                             & 67.51081  & 3.94835  & 0.023240835 & 8.726953829 & 8.3 & --\\
        \sceneloc{\texttt{ang20191023t151141}\\Carlsbad, New Mexico, USA } & Oil and natural gas extraction \chiv{}               & 66.033415 & 0.800214 & 1.090617177 & 8.608585024 & 7.2 & --\\
        \bottomrule
    \end{tabular}
    
    Zenith -- Mean Scene Solar Zenith Angle; Water -- Mean Scene Column Water Vapor;\\Ground -- Mean Scene Ground Elevation; Sensor -- Mean Scene Sensor Altitude
\end{table*}

\subsection{Benchmark Dataset}
\label{sec:methods:dataset}
A collection of  Airborne Visible/Infrared Imaging Spectrometer Next Generation (AVIRIS-NG; \citealp{Hamlin2011a}) scenes was selected for evaluating effects of unit enhancement spectra on \chiv{} and \coii{} retrieval. 
AVIRIS-NG measures reflected solar radiance at approximately 5~nm spectral resolution from 360~nm to 2500~nm, producing 425 spectral bands. 
Mounted on an airborne platform, AVIRIS-NG flights are separated into scenes of near-linear flight path segments that observe tens to hundreds of square kilometers depending on aircraft altitude and flight segment length. 

AVIRIS-NG scenes selected for the benchmark dataset spanned 2015 to 2019, and are summarized in Table~\ref{tab:scene_table}. 
Scenes were chosen to represent ranges in geometric and atmospheric parameters: solar zenith angle, sensor altitude, ground elevation, and column water vapor.
Scenes were required to contain one or more \chiv{} or \coii{} plumes to be selected for the dataset. 
Solar zenith angle, sensor altitude, and ground elevation within each scene were calculated from per-pixel scene metadata. 
Column water vapor, representing vertically integrated ground-to-space water vapor expressed in units of cm of condensed water, was derived from the standard AVIRIS-NG three-phase water product \citep{Thompson2015a}. 
Mean per-scene values were calculated for all four parameters.

Benchmark scenes contain a variety of \chiv{} point sources, including oil and gas infrastructure/wells, landfills, and waste management. 
Scenes containing \coii{} point sources are exclusively power plants.
A number of scenes that were selected have been included in previous analyses, which are referenced in the last column of Table~\ref{tab:scene_table} when applicable.
Seven of the scenes contained one or more \chiv{} plumes.
Four scenes contained one or more \coii{} plumes. 
The solar zenith angles represented by this dataset ranged from 15 to 67.5~degrees; at 67.5~degrees, the downwelling solar pathlength is over 2.6~times longer than a vertical path.
Mean column water vapor ranges from 0.8~cm to 5.5~cm across the selected scenes. 
Mean sensor altitude (which, along with ground elevation, determines spatial resolution) varies from 2.5 to 8.7~km.
Mean ground elevation represented within the selected scenes ranges from 10~m to 1.6~km above sea level. 
The benchmark dataset scenes are publicly available from the Jet Propulsion Laboratory's AVIRIS-NG web site.\footnote{\url{https://avirisng.jpl.nasa.gov/benchmark_methane_carbon_dioxide.html}} 

\subsection{Generic Unit Enhancement Spectrum Generation}
\label{sec:methods:generic}
Generic unit enhancement spectra were derived from the spectral signature of \chiv{} or \coii{} absorption as described by \citet{Thompson2015}. 
Absorption coefficients from HITRAN 2012 with water vapor continuum model MT\_CKD 2.5 were used as inputs to the Reference Forward Model \citep{Mlawer2012,Rothman2013,Dudhia2017}.
A 1~m layer of 1~ppm of \chiv{} or \coii{} at standard temperature and pressure was assumed, with one-directional transmittance through the enhancement. 
The high-resolution absorption coefficient spectrum, with units of (\ppmm)\textsuperscript{-1}, was convolved to AVIRIS-NG bands using calibration metadata for each scene.
Convolution assumed that each band had Gaussian spectral response defined by calibrated band centers and widths. This generic unit enhancement spectrum generation approach used by previous \chiv{} retrievals assumes Voigt line shapes for \chiv{} and \coii{} absorption, although once convolved to the 5 nm spectral resolution of AVIRIS-NG, differences between this approach and improved non-Voigt shapes available in HITRAN 2016 \citep{Gordon2017} should be negligible. 

\subsection{Scene-Specific Unit Enhancement Spectrum Generation}
\label{sec:methods:simulation}
For each scene, at-sensor radiance was simulated using MODTRAN6 \citep{Berk2014}. 
In contrast to the generic unit enhancement spectrum approach that uses a one-directional path, \chiv{} or \coii{} was enhanced within a layer above the ground, and absorption and scattering were calculated on both the downwelling and upwelling paths. 
Background concentrations of \chiv{} and \coii{} were set to 1.85~ppm and 410.0~ppm, respectively, and used to scale the entire column of each greenhouse gas.
\chiv{} and \coii{} enhancements were added to a uniform layer from 0 to 500~m above the ground. 
For example, a \chiv{} enhancement of 1000~\ppmm{} would be simulated as a 2~ppm increase above background, with the concentration within the 500~m layer equaling 3.85~ppm. 
Solar zenith angle, ground elevation, sensor altitude, and day of year used in MODTRAN simulations were extracted from scene metadata, described in section~\ref{sec:methods:dataset}. 
Simulations assumed a rural aerosol profile with 23~km visibility. 
The DISORT model was used for multiple scattering simulation.  
For benchmark scenes from the USA, a mid-latitude summer atmospheric model profile was assumed. 
For benchmark scenes from India, a tropical atmospheric model profile was assumed. 
Simulations were run at a spectral resolution of 0.1~cm\textsuperscript{-1}. 
At the minimum simulated wavelength of 1410~nm, this wavenumber resolution corresponds to a spectral wavelength resolution of approximately 0.02~nm.
At the maximum simulated wavelength of 2520~nm, this wavenumber resolution corresponds to a spectral wavelength resolution of 0.06~nm. This spectral resolution is finer than is necessary to simulate AVIRIS-NG radiance spectra, but was used to be consistent with simulations described in section~\ref{sec:methods:lut}. 
A range of simulated concentration enhancements was required for each parameter set.
\chiv{} concentration enhancement above background level was simulated at 0~\ppmm{} and values doubling from 1000~\ppmm{} to 64,000~\ppmm{} (\textit{i.e.}, 0, 1000, 2000, 4000, 8000, etc.). 
\coii{} concentration enhancement above background was simulated at 0~\ppmm{} and values doubling from 20,000~\ppmm{} to 1,280,000~\ppmm{}. 
Surface reflectance was set at 100\%, which removes dependence of the modeled radiance spectrum on surface reflectance such that extinction occurs only through atmospheric absorption and scattering. 
The at-sensor radiance from the MODTRAN simulations was convolved to AVIRIS-NG bands using the calibration information provided with flightline data.
The unit enhancement value for each band was determined by the regression of the natural log of the band's simulated radiance across the range of gas enhancements against the enhancement concentration-pathlength values.
The slope of the regression line, which is the unit enhancement value of each band, has units of (\ppmm)\textsuperscript{-1}.

\subsection{Portable Spectrum Generation Technique}
\label{sec:methods:lut}
Full simulation of MODTRAN-based, scene-specific unit enhancement spectra requires high computational expense for each spectrum generated (see section~\ref{sec:results:lut} for quantitative discussion), as well as licensing for radiative transfer software. 
We introduce a lookup table and interpolation approach for generating unit enhancement spectra for a range of expected parameters.
This lookup table effectively allows for amortizing the upfront computational expense of radiative transfer simulations across future unit enhancement spectrum queries, with a fixed overhead for storage of radiative transfer results.

Radiative transfer simulations using MODTRAN6 were performed at regular intervals on the five-dimensional grid of atmospheric and geometric parameters. 
We selected the discretization level for parameters within the generated lookup table to capture the trends and bounds expected for each parameter.
Specifically, solar zenith angle was sampled at 5\nobreakdash-degree increments between 0~degrees and 80~degrees. 
Ground altitude was sampled at 0~km, 0.5~km, 1~km, 2~km, and 3~km.
Sensor altitude was sampled at 1~km, 2~km, 4~km, 10~km, 20~km, and 120~km. 
The top of the highest atmospheric layer modeled in MODTRAN is 120~km, and this equivalent sensor altitude setting should accommodate future satellite observations. 
Column water vapor was sampled at 1~cm increments between 0~cm and 6~cm.
Separate lookup tables were generated for \chiv{} and \coii{}. 
Background and enhancement concentrations of \chiv{} and \coii{} matched those used in the direct simulation experiments of section~\ref{sec:methods:simulation}.
A mid-latitude summer atmosphere was assumed for all radiative transfer simulations that form the lookup tables, and DISORT multiple scattering was used.
Simulations were run at a spectral resolution of 0.1~cm\textsuperscript{-1} to allow for convolution to hypothetical instruments with finer spectral resolutions than AVIRIS-NG. 
The resulting sampling grid for the lookup table included 28,560 sample spectra for each gas.

To produce a unit enhancement spectrum, we linearly interpolated values from the lookup table at the coordinates of desired atmospheric and geometric parameters. 
Linear interpolation between the discrete parameter samples provided approximation of the radiative transfer result at each wavelength. Interpolated unit enhancement spectra were created for each of the benchmark dataset scenes. Our interpolation code bound parameter values to within the sample limits to prevent extrapolation with one exception: sensor altitudes beyond 120~km are allowable but assumed to be equivalent to 120~km. 
Using interpolated radiance spectra, a unit enhancement spectrum was derived following the same process for convolution to the calibrated instrument band-response function and regression as in section~\ref{sec:methods:simulation}.

\subsection{Gas Enhancement Retrieval and Integrated Mass Enhancement}
\label{sec:methods:ime}
Concentration-pathlength enhancements were retrieved from AVIRIS-NG radiance data using the sparse matched filter with albedo correction described by \citet{Foote2020} that optimizes for the rarity of enhancement and corrects for varying surface albedo.\footnote{Implementation available: \url{https://github.com/markusfoote/mag1c}} 
The optimization within the sparse matched filter retrieval method reflects the expectation that \chiv{} or \coii{} enhancements are a rare occurrence within a scene by using the $\ell_1$~sparsity prior. 
The matched filter was applied to each benchmark dataset scene using the generic, scene-specific, and interpolated unit enhancement spectra. Bands with wavelengths from 2122 to 2488~nm were used for \chiv{} retrievals, and from 1928 to 2200~nm for \coii{} retrievals.

\chiv{} and \coii{} plumes were selected in each scene, with a wide range in point-source emitter strength and plume dimensions across the dataset. 
The integrated mass enhancement (IME) of each plume was used as the basis for comparison between scene-specific and generic unit enhancement spectra, and between scene-specific and interpolated unit enhancement spectra. IME was calculated using the manually selected plume extent. This plume segmentation included pixels considered to be within both the generic and scene-specific retrieval images; thus, any differences in plume extent were not reflected in analyses.
The pixel-wise retrieved concentration-pathlength was converted to the associated mass of the selected gas using the image resolution and molecular weight of the selected gas \citep{Thompson2016}. 
The per-pixel mass was then accumulated among all pixels within the plume area, which remained constant across retrievals, to produce the IME value according to the equation
\begin{equation}
    \operatorname{IME} = 
        k\sum_{i=0}^N \alpha_i s_i
\end{equation}
where $i\in[0,N]$ are the pixels selected through manual plume segmentation, $s$ is the surface area of a pixel, and $\alpha_i$ is the concentration-pathlength enhancement within each pixel.
The constant $k$ represents the conversion factor from concentration-pathlength (in \ppmm) to mass of the target gas (typically in kilograms); we derive it as the molecular mass of the gas divided by the volume of one mole of an ideal gas at standard temperature and pressure with appropriate unit conversions. 
For example, for \chiv{}, $k=\sfrac{16.043\si{g}}{\si{mol}}\!\cdot\!\sfrac{1\,\si{mol}}{0.0224\si{m^3}}\!\cdot\!\sfrac{1}{10^6\si{ppm}}=0.716\!\times\!10^{-6}\,\sfrac{\si{g}}{\si{ppm\,m^3}}$. 
For \coii{} retrievals, the molecular mass is replaced with 44.009~\sfrac{g}{mol}; otherwise the equation remains the same.  

The next step in flux estimation is to multiply IME by wind speed and divide by plume length,
\begin{equation}
    Q = \operatorname{IME} \frac{U}{l}
    \label{eq:flux}
\end{equation}
where $Q$ is flux, $U$ is wind speed, and $l$ is plume length \citep{Frankenberg2016}. 
Percent difference in IME will indicate percent difference in $Q$ if wind speed and plume length are assumed to not change between generic spectrum and scene-specific spectrum retrievals.

Flux was calculated for the three power plant \coii{} plumes that had corresponding hourly regulatory emissions monitoring data \citep{EPA}. 
These three plumes were acquired in northwestern New Mexico, USA, with one 2015 scene containing two plumes and one 2019 scene containing a single plume (Table~\ref{tab:scene_table}).
\coii{} plumes from scenes \texttt{ang20180415t045439} and \texttt{ang20191004t221515} were excluded from flux comparison due to the lack of available regulatory emissions data or insufficient information to confidently assign the observed plumes to the recorded emissions data from individual generating units.
To estimate flux and account for uncertainty due to wind speed, 10~m wind speed observations from Four Corners Regional Airport were used \citep{Horel2002}. This station is approximately 20--23~km from the imaged plumes. Mean and standard deviation of observations within $\pm$2~hours of the AVIRIS-NG flightline were calculated. In 2019, wind speed was reported at a 5\nobreakdash-minute interval; however, in 2015, only three wind speed observations were reported within $\pm$2~hours of scene acquisition.
Uncertainty of flux estimates was derived from the standard deviation in wind speed and did not consider variation in assessed plume length or IME. 

\subsection{Parameter Sensitivity}
\label{sec:methods:sens}
For each scene and greenhouse gas, eight additional retrievals were performed with interpolated unit enhancement spectra that perturbed individual parameters to their extrema within the benchmark dataset (\textit{i.e.}, minimum and maximum for solar zenith angle, ground elevation, sensor altitude, and column water vapor). One parameter was varied while the remaining three parameters were fixed at their scene-specific values.
IME calculations were performed on these retrievals with the same plume segmentation masks and protocol as described in section~\ref{sec:methods:ime}. The varied parameter was used to calculate percent change in IME (relative to the same plume with `correct' scene-specific values) with change in that parameter for each plume. Sensitivity to each parameter was determined for nineteen \chiv{} plumes and seven \coii{} plumes.

\section{Results}
Results from unit enhancement spectra generation are described in section~\ref{sec:results:uas}, followed by the application of these unit enhancement spectra to concentration-pathlength retrievals in section~\ref{sec:results:retrievals}, including analysis of the effect of the unit enhancement spectra on IME.
Section~\ref{sec:results:flux} extends IME measurements to flux approximations for three power plant \coii{} plumes and compares the flux approximations to reported values.
Section~\ref{sec:results:lut} describes the errors that are introduced by using a lookup table approximation for portable scene-specific unit enhancement spectrum generation. Finally, section~\ref{sec:results:sens} outlines the relative importance of each parameter for accurate IME calculation.

\subsection{Unit Enhancement Spectra Generation}
\label{sec:results:uas}

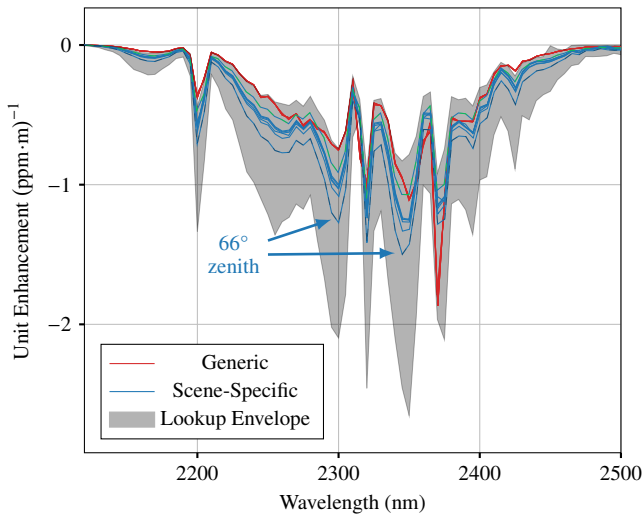
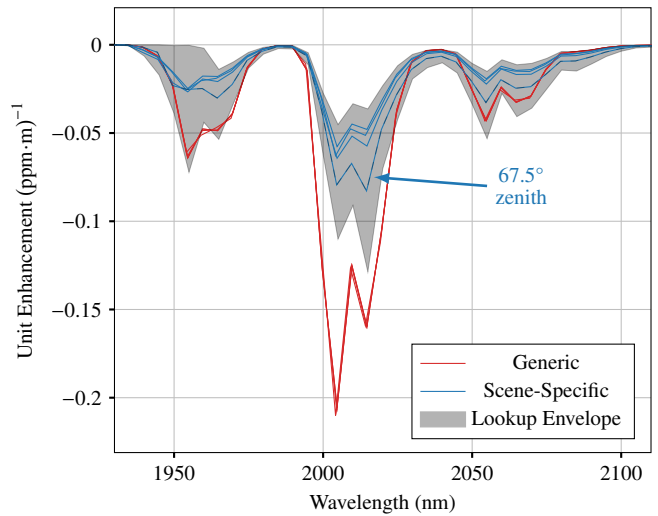
\begin{figure*}[tp]
    \centering
    \definecolor{myinterp}{rgb}{0.4546,0.6530,0.4208}   
    \definecolor{mygeneric}{HTML}{d62728}
    \definecolor{myspecific}{HTML}{1f77b4}
    \colorlet{shadecolor}{black!100}
    \newcommand{\shadeopacity}{0.3}
    \newlength{\spectrafigureheight}
    \setlength{\spectrafigureheight}{7.5cm}
    \begin{subfigure}[b]{.5\textwidth}
        \centering
        \begin{tikzpicture} 
        \footnotesize
        \begin{axis}[
            width=0.95\textwidth,
            height=\spectrafigureheight,
            xlabel={Wavelength (nm)}, 
            ylabel={Unit Enhancement $\left(\textrm{ppm}\!\cdot\!\textrm{m}\right)^{-1}$}, 
            grid=major,
            no markers,
            xmin=2120,
            xmax=2500,
            legend pos=south west,
            x tick label style={/pgf/number format/.cd,%
                scaled x ticks = false,
                set thousands separator={},
                fixed,%
                },%
            tick align=outside,
            ytick pos=left,
            xtick pos=lower,
            major tick length=2pt,
            major tick style = {thick, black},
            ]
            \foreach \hitran in {ang20150420t181345, ang20160211t075004, ang20170616t212046, ang20170618t193955, ang20170906t210217, ang20180927t195028, ang20191023t151141} {
                \addplot[color=mygeneric,forget plot] table [x index=1,y index=2] {figures/spectra/hitran/benchmark_\hitran_ch4_unit.txt}; 
            }%
            \addlegendimage{line legend, mygeneric}
            \addlegendentry{Generic}
            \foreach \modtran in {ang20150420t181345, ang20160211t075004, ang20170616t212046, ang20170618t193955, ang20170906t210217, ang20180927t195028, ang20191023t151141} {
                \addplot[color=myspecific, forget plot] table [x index=1,y index=2] {figures/spectra/modtran/\modtran_ch4_ua.txt};
            }%
                \addplot[color=black, opacity=0.3, dashed, forget plot] table [x index=1,y index=2] {figures/spectra/modtran/ang20191023t151141_ch4_ua.txt}; 
                \addplot[color=green, opacity=0.5, dashed, forget plot] table [x index=1,y index=2] {figures/spectra/modtran/ang20150420t181345_ch4_ua.txt};
            \addlegendimage{line legend, myspecific}
            \addlegendentry{Scene-Specific}
            \addplot[name path=low,
                     forget plot,
                     no markers,
                     line width=0pt,
                     shadecolor,
                     opacity=\shadeopacity,
                    ] table [x=wave,y=ch4_low] {figures/spectra/lut_bounds/uas_envelope_spectra.txt};
            \addplot[name path=high,
                     forget plot,
                     no markers,
                     line width=0pt,
                     shadecolor,
                     opacity=\shadeopacity,
                    ] table [x=wave,y=ch4_high] {figures/spectra/lut_bounds/uas_envelope_spectra.txt};
            \addplot [shadecolor, opacity=\shadeopacity] fill between[ 
                     of = low and high, 
                     ];
            \addlegendentry{Lookup Envelope}
            \draw[myspecific, -latex, line width = 1pt] (axis cs:2250,-1.4) -- (axis cs:2295,-1.25);
            \draw[myspecific, -latex, line width = 1pt] (axis cs:2250,-1.5) -- (axis cs:2340,-1.49) node[anchor = east, pos=0, font=\footnotesize] {\shortstack{66\textdegree\\zenith}};
        \end{axis}
        \end{tikzpicture}
        \caption{\chiv{} unit enhancement spectra.}
        \label{fig:spectra:ch4} 
    \end{subfigure}%
    \begin{subfigure}[b]{.5\textwidth}
        \centering
        \begin{tikzpicture} 
        \footnotesize
        \begin{axis}[
            width=0.95\textwidth,
            height=\spectrafigureheight,
            xlabel={Wavelength (nm)}, 
            ylabel={Unit Enhancement $\left(\textrm{ppm}\!\cdot\!\textrm{m}\right)^{-1}$}, 
            grid=major,
            no markers,
            xmin=1930, 
            xmax=2110, 
            legend pos=south east,
            yticklabel style={
                /pgf/number format/fixed,
                /pgf/number format/precision=2,
            },
            x tick label style={/pgf/number format/.cd,%
                scaled x ticks = false,
                set thousands separator={},
                fixed,%
                },%
            tick align=outside,
            ytick pos=left,
            xtick pos=lower,
            major tick length=2pt,
            major tick style = {thick, black},
            ]
            \foreach \hitran in {ang20150420t181345, ang20180415t045439, ang20190621t200919, ang20191004t221515} {
                \addplot[color=mygeneric,forget plot] table [x index=1,y index=2] {figures/spectra/hitran/benchmark_\hitran_co2_unit.txt}; 
            }%
            \addlegendimage{line legend, mygeneric}
            \addlegendentry{Generic}
            \foreach \modtran in {ang20150420t181345, ang20180415t045439, ang20190621t200919, ang20191004t221515} {
                \addplot[color=myspecific, forget plot] table [x index=1,y index=2] {figures/spectra/modtran/\modtran_co2_ua.txt};
            }%
            \addlegendimage{line legend, myspecific}
            \addlegendentry{Scene-Specific}
            \addplot[color=black, opacity=0.3, dashed, forget plot] table [x index=1,y index=2] {figures/spectra/modtran/ang20191004t221515_co2_ua.txt}; 
            \addplot[name path=co2low,
                     forget plot,
                     no markers,
                     line width=0pt,
                     shadecolor,
                     opacity=\shadeopacity,
                    ] table [x=wave,y=co2_low] {figures/spectra/lut_bounds/uas_envelope_spectra.txt};
            \addplot[name path=co2high,
                     forget plot,
                     no markers,
                     line width=0pt,
                     shadecolor,
                     opacity=\shadeopacity,
                    ] table [x=wave,y=co2_high] {figures/spectra/lut_bounds/uas_envelope_spectra.txt};
            \addplot [shadecolor, opacity=\shadeopacity] fill between[ 
                     of = co2low and co2high, 
                     ];
            \addlegendentry{Lookup Envelope}
            \draw[myspecific, -latex, line width = 1pt] (axis cs:2055,-0.08) -- (axis cs:2017,-0.075)
        node[anchor = west, pos=0, font=\footnotesize] {\shortstack{67.5\textdegree\\zenith}};
        \end{axis}
        \end{tikzpicture}
        \caption{\coii{} unit enhancement spectra.}
        \label{fig:spectra:co2}
    \end{subfigure}
    \caption{Unit enhancement spectra for the collection of ten scenes in the benchmark dataset, showing spectra for (a) methane and (b) carbon dioxide. Generic unit enhancement spectra are shown in red. Results from the scene-specific method (blue) are more varied, as they take into account geometric and atmospheric parameters. 
    These specific spectra lines are discussed in section \ref{sec:results:uas}. 
    The 66\textdegree{} and 67.5\textdegree{} labels indicate scenes with the largest solar zenith angles.
    The gray shaded regions represent the range of results from the portable unit enhancement generation method using a lookup table, discussed in section \ref{sec:results:lut}.
    }
    \label{fig:spectra}
\end{figure*}
Using the geometric and atmospheric parameters from the benchmark dataset, scene-specific unit enhancement spectra were generated for \chiv{} and \coii{}. 
These spectra are compared to generic unit enhancement spectra in Figure~\ref{fig:spectra}.
The generic unit enhancement spectra for both \chiv{} and \coii{} differed only in the convolution to sensor bands resulting from variations in instrument calibration; thus, these spectra appeared as one (slightly bolder) line for both \chiv{} and \coii{}.
Unit enhancement spectra from the scene-specific generation technique showed much greater variance due to differences in geometric and atmospheric parameters between scenes. 
The gray lookup table envelope, which represents a wider range in geometric and atmospheric parameters, creates a region that contains all scene-specific unit enhancement spectra within allowable parameter ranges. 

For \chiv{}, the magnitudes of the generic and scene-specific unit enhancement spectra appear similar (Figure~\ref{fig:spectra:ch4}).
One-way transmittance used by the generic unit enhancement spectrum generation technique should result in underestimation of the magnitude of the enhancement spectrum relative to two-way transmittance through a uniform layer, whereas not including the atmospheric background concentration of \chiv{} should result in overestimation of the magnitude of the enhancement spectrum. 
These two effects largely cancel out for \chiv{}, particularly for low pathlength conditions (small solar zenith angle and low sensor altitude). 
For example, the spectrum for scene \texttt{ang20150420t181345} is closest to the generic unit enhancement spectrum -- in Figure~\ref{fig:spectra:ch4} it is annotated with a green dashed highlight. 
This scene had a relatively low solar zenith angle, but had the highest ground elevation and smallest difference between ground elevation and sensor altitude (Table~\ref{tab:scene_table}), resulting in shorter pathlengths. 
Longer pathlengths produced by high solar zenith angles, high sensor altitudes, and low ground elevations produce higher magnitude scene-specific enhancement spectra, which will produce lower retrieved concentration-pathlengths versus the generic \chiv{} unit enhancement spectrum.    
The largest magnitude scene-specific \chiv{} unit enhancement spectrum occurs at a large solar zenith angle, as indicated by the label with a 66\textdegree{} solar zenith angle for scene \texttt{ang20191023t151141} in Figure~\ref{fig:spectra:ch4}. 

The combined effects of atmospheric water vapor absorption, pathlength, and background \chiv{} resulted in important differences in shape between the generic \chiv{} unit enhancement spectrum and the scene-specific spectra. The generic spectrum has its maximum magnitude at an absorption coefficient peak near 2370~nm. Scene-specific spectra have lower magnitudes at this wavelength (Figure~\ref{fig:spectra:ch4}). When water vapor was omitted from simulations, peak magnitude still occurred near 2370~nm for all solar zenith angles less than 30\textdegree{}. At larger solar zenith angles, change in radiance at 2370~nm was reduced due to absorption being closer to saturation, resulting in mixture of peak magnitudes (either 2345~nm or 2370~nm) up to a solar zenith angle of 65\textdegree{}, with longer pathlengths shifting peak magnitude to the shorter wavelength. Above 65\textdegree{}, all dry atmosphere simulations peaked near 2345~nm. Once water vapor was added to scene-specific simulations, all had lower magnitudes at 2370~nm due to stronger water vapor absorption in this portion of the \chiv{} absorption spectrum. Although most of the scene-specific spectra have peak magnitudes near 2345~nm or 2350~nm, the combination of geometric and atmospheric parameters for scene \texttt{ang20150420t181345} produces a scene-specific unit enhancement spectrum that peaks at near 2320~nm (green dashed line in Figure~\ref{fig:spectra:ch4}). 

Unit enhancement values for \coii{} were much lower than those for \chiv{}, due to relatively weaker absorption by \coii{}. Scene-specific unit enhancement spectra for \coii{} all had a lower magnitude than the generic unit enhancement spectrum (Figure~\ref{fig:spectra:co2}). 
This result indicates that the relatively high background concentration of \coii{} in the atmosphere is dominant over the effect of one-way transmittance simulated by the generic unit enhancement spectrum, producing a reduced magnitude of absorption in the scene-specific spectra. 
As with \chiv{}, scenes with the largest solar zenith angles had the largest magnitude \coii{} unit enhancement spectra, as indicated by the labeled 67.5\textdegree{} solar zenith angle representing scene \texttt{ang20191004t221515} in Figure~\ref{fig:spectra:co2}. 
The \coii{} unit enhancement spectra also show subtle differences in shape due to differences in background water vapor absorption. 
The generic spectrum peaks near 2005~nm, but the spectrum for \texttt{ang20191004t221515} peaks at a slightly longer wavelength near 2015~nm due to higher water vapor and solar zenith angle.      

Taken collectively, the generic approach to modeling unit enhancement spectra would seem to slightly overestimate \chiv{} concentration-pathlength, on average, due to a lower magnitude enhancement spectra relative to many scene-specific unit enhancement spectra. 
However, this effect is complicated by the pronounced differences in spectral shape between \chiv{} generic and scene-specific spectra, with the generic spectrum more weighted toward longer wavelengths and the scene-specific spectra more weighted toward shorter wavelengths due to water vapor absorption and pathlength effects. 
In contrast, the generic approach to modeling unit enhancement spectra should greatly underestimate \coii{} concentration-pathlength due to a much higher magnitude absorption relative to scene-specific unit enhancement spectra. The differences in the magnitudes of generic and scene-specific \coii{} spectra would appear to be much more important than the more subtle differences in spectral shape.   

\subsection{Enhancement Retrievals and IME}
\label{sec:results:retrievals}
\newlength{\comparisonfigheight}
\setlength{\comparisonfigheight}{5in}
\begin{figure*}[t!]
    \centering
    \begin{subfigure}[b]{0.5\textwidth}
    \centering%
        \includegraphics[height=\comparisonfigheight,trim={140pt 90pt 65pt 85pt},clip]{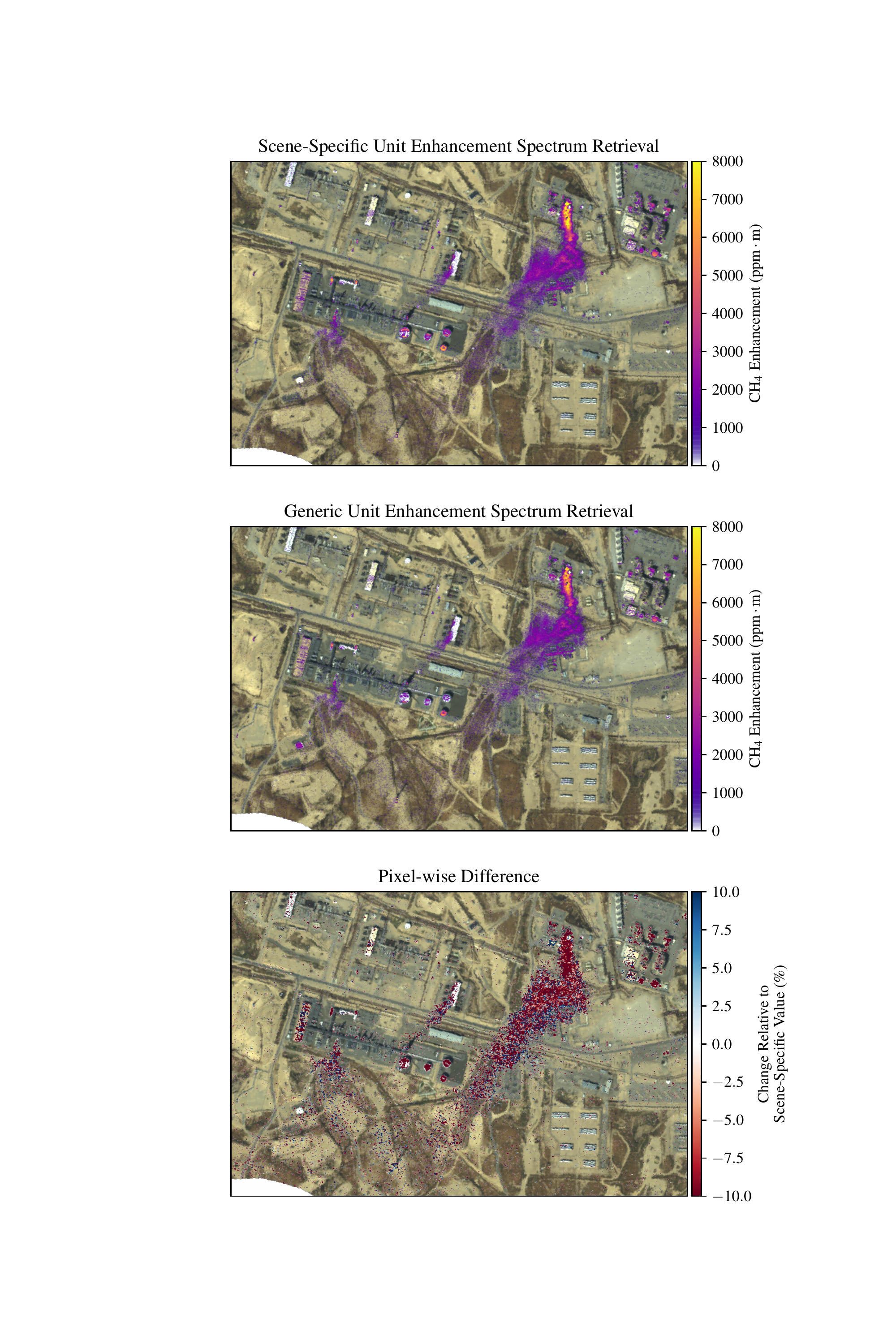}
        \caption{\chiv{} enhancement retrievals for \texttt{ang20170906t210217}.}
        \label{fig:retrieval_comparison:ch4}
    \end{subfigure}%
    \begin{subfigure}[b]{0.5\textwidth}
    \centering%
        \includegraphics[height=\comparisonfigheight,trim={120pt 90pt 40pt 85pt},clip]{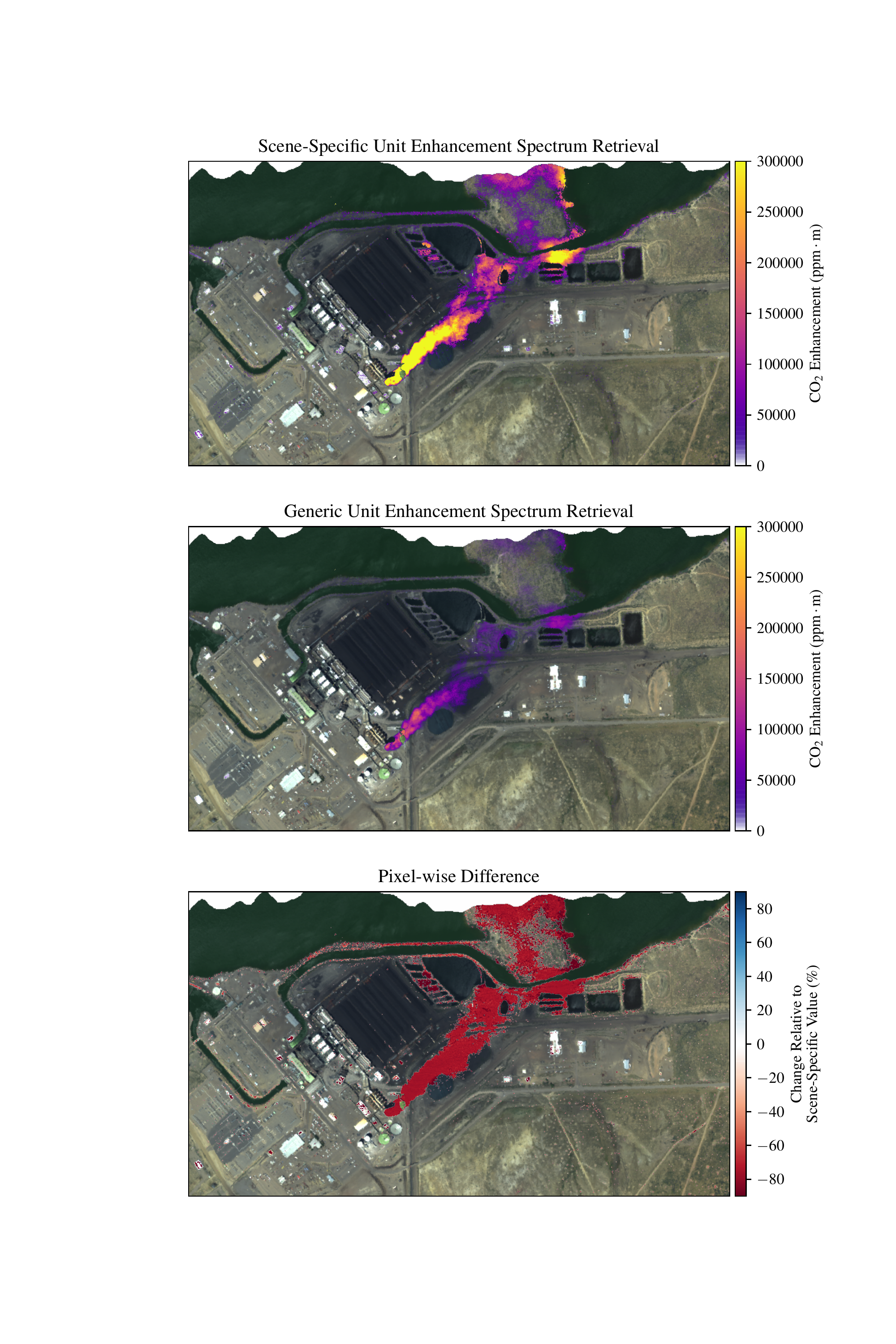}
        \caption{\coii{} enhancement retrievals for \texttt{ang20190621t200919}.}
        \label{fig:retrieval_comparison:co2}
    \end{subfigure}%
    \caption{Concentration-pathlength enhancement retrievals with the scene-specific (top) and generic (middle) unit enhancement spectra for two scenes from the benchmark dataset. 
    The top two panes in each column have a common colormap for qualitative comparison for each scene. The basemap data are AVIRIS-NG true color bands. 
    White artifacts at the edges of the images are edges of the scene data. 
    The bottom panes show a pixel-wise calculated difference percentage. 
    Pixels with a higher concentration-pathlength value in the scene-specific method are highlighted in red shades, whereas pixels with lower concentration-pathlength value in the scene-specific retrieval are highlighted in blue shades.}
    \label{fig:retrieval_comparison}
\end{figure*}
We performed gas concentration-pathlength retrievals for all the benchmark scenes using generic and scene-specific unit enhancement spectra in the sparse matched filter algorithm described by \citet{Foote2020}.
Figure~\ref{fig:retrieval_comparison:ch4} shows a representative retrieval for \chiv{} over natural gas infrastructure in scene \texttt{ang20170906t210217}.
The retrieval that used the scene-specific unit enhancement spectrum, displayed in the top pane, generally provided higher enhancement values than the retrieval that used the generic unit enhancement spectrum, shown in the middle pane. 
The difference in enhancement values was calculated for each pixel individually, and is shown in the bottom pane of Figure~\ref{fig:retrieval_comparison:ch4}. 
This per-pixel difference is displayed as a percentage relative to the scene-specific retrieval value, which is generally higher; thus, these percentages are lower than if calculated as the change with respect to the generic retrieval values.

A representative sparse matched filter retrieval for a power plant \coii{} plume from scene \texttt{ang20190621t200919} is shown in Figure~\ref{fig:retrieval_comparison:co2}. 
The enhancement retrievals using scene-specific and generic unit enhancement spectra in the top and middle panes, respectively, show a stark difference in the enhancements.
The bottom pane indicates per-pixel differences of greater than 60\% lower for the generic enhancement retrieval.
Phrased differently, retrievals that used the generic unit enhancement spectra captured less than 40\% of the \coii{} enhancement that was observed with a scene-specific unit enhancement spectrum. 
These changes were more quantitatively measured using the IME values for plumes in each scene.

\begin{table*}[t!]
    \centering
    \caption{Integrated mass enhancements (IME) for each plume within the 10-scene benchmark dataset using sparse matched filter retrievals using the generic and the scene-specific unit enhancement spectra. 
    Plumes within each scene are presented successively, with the gray row highlight modulating with changing scenes.
    The manually annotated plume regions are the same for all IME calculations.
    The change in mass between unit enhancement spectra, as a percentage of the scene-specific IME value, is calculated in the last column.
    IME and plume measurement values include three significant figures; however, error values were calculated before rounding and describe more specific changes in the IMEs.
    }
    \label{tab:ime_change}
    \setlength{\tabcolsep}{4pt}
    \renewcommand{\arraystretch}{1.2}
    \footnotesize
    \newcolumntype{Y}{S[round-mode=figures,round-precision=3, group-minimum-digits=7,table-format = 5.2]}
    \newcolumntype{Z}{S[round-mode=places,round-precision=1, group-minimum-digits=7,table-format = 2.1]}
    \newcolumntype{P}{S[round-mode=figures,round-precision=3, group-minimum-digits=7,table-format = 6.0]}
    \begin{filecontents*}{figures/data/ime_change_generic_modtran.csv}
scene, gas, hitransparse, hitranjpl, modtransparse, modtranjpl, diffpctsparse, diffpctjpl, plumelength, plumearea
ang20150420t181345, \chiv, 62.39440864165059, 40.37582717440424, 69.26082648512565, 45.84925838394599, -9.91385490461867, -11.937883844721583, 501.6, 36012.020000000004
ang20150420t181345, \coii, 645.3471239499237, 356.5158155288756, 2529.1269140168693, 1340.850676951014, -74.4834029335066, -73.41122157319086, 243.10000000000002, 14628.900000000001
ang20150420t181345, \coii, 7039.651645759102, 2368.5534598455006, 25981.12096590193, 8477.996105867247, -72.9047424281729, -72.06234314962319, 790.9000000000001, 142875.59000000003
ang20160211t075004, \chiv, 4.540893655506537, 6.134298739493684, 4.334067826950455, 5.768337854332591, 4.772094872857792, 6.344303929531804, 259.2, 14696.64
ang20160211t075004, \chiv, 13.554539689799748, 17.585004098566316, 14.472284943448134, 17.524628196128937, -6.341398453903893, 0.34452030457750465, 356.4, 27359.37
ang20160211t075004, \chiv, 165.40461745306501, 173.87929999605961, 128.5242812092816, 145.29476178061952, 28.695228556641034, 19.673481593644695, 1506.6, 323850.96
ang20160211t075004, \chiv, 11.368669219356773, 13.660358255896258, 11.02293444380221, 14.077015876200443, 3.1365039619641113, -2.95984336430717, 234.89999999999998, 12925.17
ang20170616t212046, \chiv, 2.0731779773760013, 2.7007641513087006, 1.8902660472972839, 2.763552825233032, 9.676517775910234, -2.2720272741316876, 110.10000000000001, 2277.0
ang20170616t212046, \chiv, 23.18629331088947, 31.767113295289228, 25.878702137619072, 36.30644959643366, -10.403956166007756, -12.502837241321235, 547.2, 47907.0
ang20170616t212046, \chiv, 2.062650602399332, 2.070605680098226, 2.2489056224981803, 2.4254551873861727, -8.282029189466316, -14.630223189996567, 122.39999999999999, 2745.0
ang20170618t193955, \chiv, 6.725910013973007, 10.236418333648455, 6.487245578445725, 10.132407105488387, 3.6789795089653943, 1.0265204218228565, 231.0, 12904.649999999998
ang20170906t210217, \chiv, 2.9041311289915925, 3.4582228216711153, 3.1820477286500086, 3.661582967772359, -8.733891611874812, -5.55388606215208, 147.21, 7867.4400000000005
ang20170906t210217, \chiv, 1.752122999949803, 1.8055807304253848, 2.126880324349354, 2.041547563150804, -17.62004754612581, -11.558233419810318, 159.18, 4581.99
ang20170906t210217, \chiv, 38.640975042067154, 34.75457888257227, 44.709285668156085, 40.43798130048249, -13.572819461105922, -14.05461458542791, 632.3100000000001, 59332.14
ang20170906t210217, \chiv, 0.25183928017379503, 0.2795641041574986, 0.3228876017028015, 0.3567825608835334, -22.004041392212436, -21.643001982723494, 73.71000000000001, 780.57
ang20180415t045439, \coii, 6466.338176029015, 8484.823362960573, 20849.241715946842, 26833.19239696514, -68.9852596841297, -68.37937418165639, 959.2, 212896.0
ang20180927t195028, \chiv, 162.7292294917562, 284.5988660608136, 158.98024775888496, 273.54454608291223, 2.35814309369871, 4.041140697629096, 1535.04, 470773.76000000007
ang20180927t195028, \chiv, 4.659678984430372, 7.325807980868155, 4.530222701064088, 7.075038049804719, 2.8576141154358026, 3.5444322602668934, 207.36, 9871.360000000002
ang20190621t200919, \coii, 9780.218778518052, 5136.618902327342, 40914.994336784235, 21161.303473717944, -76.0962479964827, -75.72635868718129, 1154.5200000000002, 211286.07000000004
ang20191004t221515, \coii, 1562.3860157890217, 1083.60696589094, 3235.616330795702, 2337.90435847992, -51.712877669745225, -53.65050063059511, 161.02, 12124.640000000003
ang20191004t221515, \coii, 1253.4924011867831, 893.6420694729466, 2414.6651378524875, 1697.659562056228, -48.08835471482426, -47.3603489506132, 230.40800000000004, 13295.770000000002
ang20191004t221515, \coii, 1907.1243526870812, 1641.0033092859746, 4172.539702891957, 3201.2657537396567, -54.29344024299476, -48.73892280361334, 403.38000000000005, 26729.320000000007
ang20191023t151141, \chiv, 54.84492419539909, 55.82212782970508, 58.62210636612049, 62.64467094569466, -6.443272691587132, -10.890859530420233, 455.76, 75582.72
ang20191023t151141, \chiv, 228.20128852018743, 259.0245923548819, 223.41834506754807, 260.06433743994324, 2.140801576161209, -0.3998030238580618, 978.48, 329184.0
ang20191023t151141, \chiv, 272.05078241592764, 402.010230340053, 289.82268291477686, 414.89048934655625, -6.1319908849491584, -3.1044960868564107, 1985.7600000000002, 582163.2000000001
ang20191023t151141, \chiv, 144.87172567319138, 189.05880573134954, 133.472053768567, 179.42457600038395, 8.540867981540732, 5.369515116449248, 702.72, 131258.88
\end{filecontents*}
\pgfplotstabletypeset[
        columns={scene,gas,plumelength,plumearea,modtransparse,hitransparse,diffpctsparse},
        sort,
        sort key=gas,
        sort cmp=string <,
        col sep=comma,
        trim cells=true,
        format=file,
        row sep=newline,
        every nth row={18}{after row=\cmidrule(lr){1-7}},
        every head row/.style={
            before row={\toprule},
            after row=\midrule},
        every last row/.style={after row=\bottomrule},
        every nth row={100[1]}{before row={\rowcolor[gray]{0.9}}},
        every nth row={100[2]}{before row={\rowcolor[gray]{0.9}}},
        every nth row={100[3]}{before row={\rowcolor[gray]{0.9}}},
        every nth row={100[4]}{before row={\rowcolor[gray]{0.9}}},
        every nth row={100[8]}{before row={\rowcolor[gray]{0.9}}},
        every nth row={100[13]}{before row={\rowcolor[gray]{0.9}}},
        every nth row={100[14]}{before row={\rowcolor[gray]{0.9}}},
        every nth row={100[19]}{before row={\rowcolor[gray]{0.9}}},
        every nth row={100[20]}{before row={\rowcolor[gray]{0.9}}},
        every nth row={100[22]}{before row={\rowcolor[gray]{0.9}}},
        columns/scene/.style={
            column name=Scene,
            string type},
        columns/gas/.style={
            column name=Gas,
            string type},
        columns/hitransparse/.style={
            column name={\textrm{Generic IME (kg)}},
            column type=Y,
            string type,
            },
        columns/modtransparse/.style={
            column name={\textrm{Scene-Specific IME (kg)}},
            column type=Y,
            string type,
            },
        columns/diffpctsparse/.style={
            column name={\textrm{Change (\%)}},
            column type=Z,
            string type
            },
        columns/diffpctjpl/.style={
            column name={\textrm{Robust MF}},
            column type=Z,
            string type
            },
        columns/modtranjpl/.style={
            column name={\textrm{Robust MF}},
            column type=Y,
            string type
            },
        columns/hitranjpl/.style={
            column name={\textrm{Robust MF}},
            column type=Y,
            string type
            },
        columns/plumelength/.style={
            column name={\textrm{Plume Length (m)}},
            column type=P,
            string type
            },
        columns/plumearea/.style={
            column name={\textrm{Plume Area (m\textsuperscript{2})}},
            column type=P,
            string type
            },
    ]{figures/data/ime_change_generic_modtran.csv}
\end{table*}

IME values resulting from retrievals using generic and scene-specific unit enhancement spectra are provided in Table~\ref{tab:ime_change}. Percent change was calculated relative to the scene-specific IME value.
The third plume within \texttt{ang20160211t\-075004} had the largest \chiv{} IME increase from scene-specific to generic retrieval, with a generic spectrum IME 28.7\% higher than the scene-specific retrieval IME.
The \chiv{} IME from the generic retrieval of the fourth plume in \texttt{ang20170906t210217} had the largest decrease of 22.0\% from the scene-specific to the generic IME. 
Other plumes in this scene also produced lower generic IME, ranging from an 8.7\% to 17.6\% decrease from scene-specific IME. 
Across all \chiv{} plumes, the generic spectrum produced IMEs that averaged 2.3\% lower than the scene-specific spectrum. Smaller plumes tended to have larger reductions in IME; in the 12 \chiv{} plumes with IME less than 50 kg, the IME from generic spectrum retrieval averaged 5.2\% lower than the scene-specific spectrum.
The varying increase and decrease of \chiv{} IME across scenes reflects the influence of the geometric and atmospheric input parameters. 

\begin{figure}[t!]
    \centering
    \includegraphics[width=1\columnwidth]{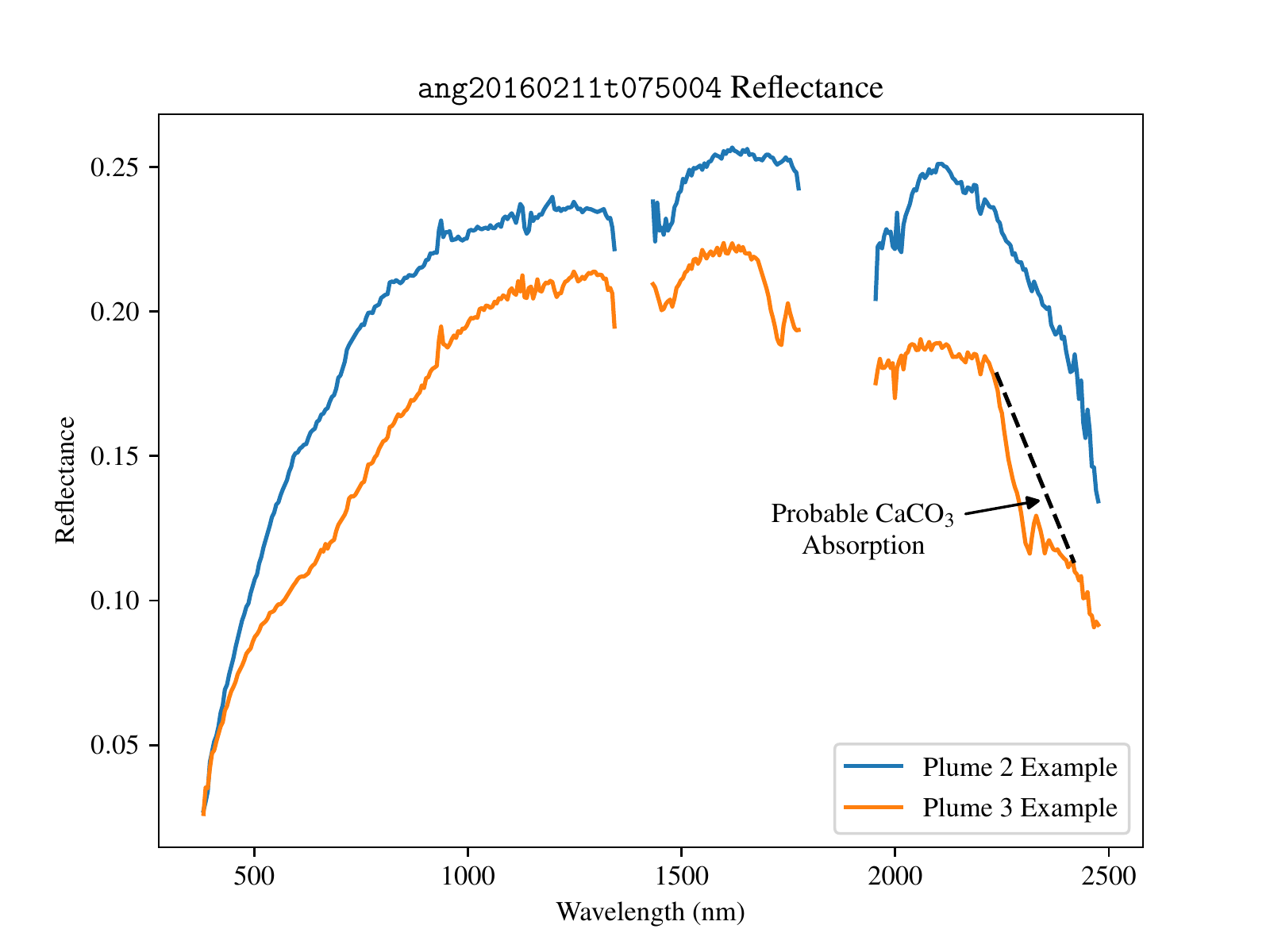}
    \caption{
    Example soil reflectance spectra for selected pixels from the second and third plumes within scene \texttt{ang20160211t075004}. 
    The third plume includes substantial regions over a landfill with surface reflectance features that affected matched filter retrieval. 
    Absorption from approximately 2200--2400 nm in the spectrum from the third plume is similar to calcium carbonate absorption.  }
    \label{fig:reflectance}
\end{figure}

Within-scene differences in IME between scene-specific and generic enhancement spectra are associated with plumes that are located over different background surfaces. Since the scene-specific enhancement spectra have subtle differences in shape compared to the generic enhancement spectra, background spectral features, and particularly shortwave infrared absorption features, can have substantial impacts on IME. 
Figure~\ref{fig:reflectance} includes two example reflectance spectra of pixels within the 2nd and 3rd plumes in scene \texttt{ang20160211t075004}, with respective IME changes of $-$6.3\% and 28.7\%. 
The orange Plume 3 reflectance spectrum demonstrates evidence of calcium carbonate absorption in the shortwave infrared, whereas the blue Plume 2 reflectance spectrum shows no such absorption. 
The generic and scene-specific spectra impart different sensitivities to surface reflectance, in part due to the weighting toward shorter wavelengths in the scene-specific spectra from water vapor absorption and pathlength effects. 
These differing sensitivities become more pronounced with distinct surface reflectance features in the shortwave infrared. \citet{Ayasse2018} noted that calcium carbonate is an important "confuser" background spectrum with the potential to produce large errors in methane retrievals.   

The IME results for \coii{} were consistently higher for enhancement retrievals that used scene-specific unit enhancement spectra, regardless of the matched filter method. 
The generic unit enhancement spectrum retrievals contained 48.1--76.1\% less \coii{} mass enhancement, which was a direct result of scene-specific unit enhancement spectra that are of a similarly lower magnitude, as observed in Figure~\ref{fig:spectra:co2}.
\coii{} plume IME values from the generic unit enhancement spectrum were approximately half those from the secne-specific spectrum for scene \texttt{ang20191004t221515}, which had a large zenith angle of 67.5\textdegree{}. In scenes with smaller solar zenith angles, plume IMEs for generic enhancement spectra were less than a third of those from scene-specific unit enhancement spectra. 
These results demonstrate the importance of both the background concentration of \coii{} and pathlength due to solar zenith angle.

\subsection{\texorpdfstring{\coii{}}{CO2} Flux Estimates}
\label{sec:results:flux}
\begin{table*}[tbh]
    \centering
    \caption{Calculated flux estimates for selected \coii{} plumes with attributable regulatory monitoring data.}
    \label{tab:flux}\vspace{-5pt}
    \setlength{\tabcolsep}{5pt}
    \renewcommand{\arraystretch}{1.3}
    \newcolumntype{F}{S[round-mode=figures,round-precision=3,table-format=6,table-number-alignment=right]@{\,\( \pm \)\,}}
    \newcolumntype{G}{S[round-mode=figures,round-precision=3,table-format=6,table-number-alignment=left]}
    \newcolumntype{K}{S[round-mode=figures,round-precision=3,table-format=4.2,table-number-alignment=right]@{\,\( \pm \)\,}}
    \newcolumntype{J}{S[round-mode=figures,round-precision=3,table-format=1.2,table-number-alignment=left]}
    \footnotesize
    \begin{tabular}{cKJFGFGS[round-mode=figures,round-precision=3,table-format=7]}
    \toprule
     & \multicolumn{2}{c}{\multirow[b]{2.5}{*}{\shortstack{Wind Speed (m/s)\\(mean$\,\pm\,$s.d.)}}}& \multicolumn{5}{c}{Flux (kg/hr)}\\
     \cmidrule(lr){4-8}
    Scene &\multicolumn{2}{c}{\phantom{A}} & \multicolumn{2}{c}{Scene-Specific} & \multicolumn{2}{c}{Generic} & {Reported} \\
    \midrule
    ang20150420t181345 & 3.9433333  & 1.94218663 & 148000 & 72800   & 37700  & 18600  & 227068.41  \\
    ang20150420t181345 & 3.9433333  & 1.94218663 & 466000 & 230000  & 126000 & 62300  & 300641.11  \\
    ang20190621t200919 & 6.63846154 & 2.48013444 & 847000 & 317000  & 202000 & 75700  & 1191224.62 \\
    \bottomrule
    \end{tabular}
\end{table*}
The large changes in \coii{} IME between generic and scene-specific spectra raise the question of which spectra would provide more accurate emission fluxes. Comparing \coii{} flux estimated from three power plant plumes to EPA emissions data revealed that flux estimates based on scene-specific spectra were much closer to reported values than those based on generic spectra (Table~\ref{tab:flux}). 
EPA flux values represent the emissions reported for the same hour as the AVIRIS-NG flight and do not indicate flux variation within that hour.
For two of the three plumes, the scene-specific spectra underestimated reported flux with a difference of more than one standard deviation. For one plume, the scene-specific spectrum overestimated reported flux but was within the one standard deviation estimate. 
In contrast, flux estimates based on generic spectra underestimate flux by a factor of 2.4\texttimes{} to 6.0\texttimes{}. 
Regulatory reported flux values were within the 95\% confidence interval for all of the flux estimates derived from scene-specific unit enhancement spectra, but were not within any of the 95\% confidence intervals from generic spectra.

\subsection{Unit Enhancement Spectrum Interpolation}
\label{sec:results:lut}
Simulation of two lookup tables for \chiv{} and \coii{} at 0.1~cm\textsuperscript{-1} resolution generated approximately 7~GB of radiance data for each gas, as described by the simulation strategy in section~\ref{sec:methods:lut}.\footnote{Data/Interpolation Implementation Available: \url{https://doi.org/10.7278/S50D-0D0H-09A6}}  Across all plumes in the benchmark dataset, root mean square IME error was 0.50\% for the interpolated \chiv{} unit enhancement spectra and 0.72\% for the interpolated \coii{} unit enhancement spectra. The latter was heavily influenced by a $-$1.9\% difference for a \coii{} plume in scene \texttt{ang20180415t045439}. 
This scene was simulated with a tropical atmospheric model for the scene-specific unit enhancement spectrum; however, the lookup table (and the resulting interpolated spectrum) used a mid-latitude summer atmospheric model. 
A special re-simulation of the appropriate parameters but replacing the assumed tropical model with a mid-latitude summer model reduced the error in interpolation to 0.15\%.

\begin{figure*}[t]
    \centering
    \includegraphics[width=\textwidth,trim={0.9in 0cm 1.25in 0.4in}, clip]{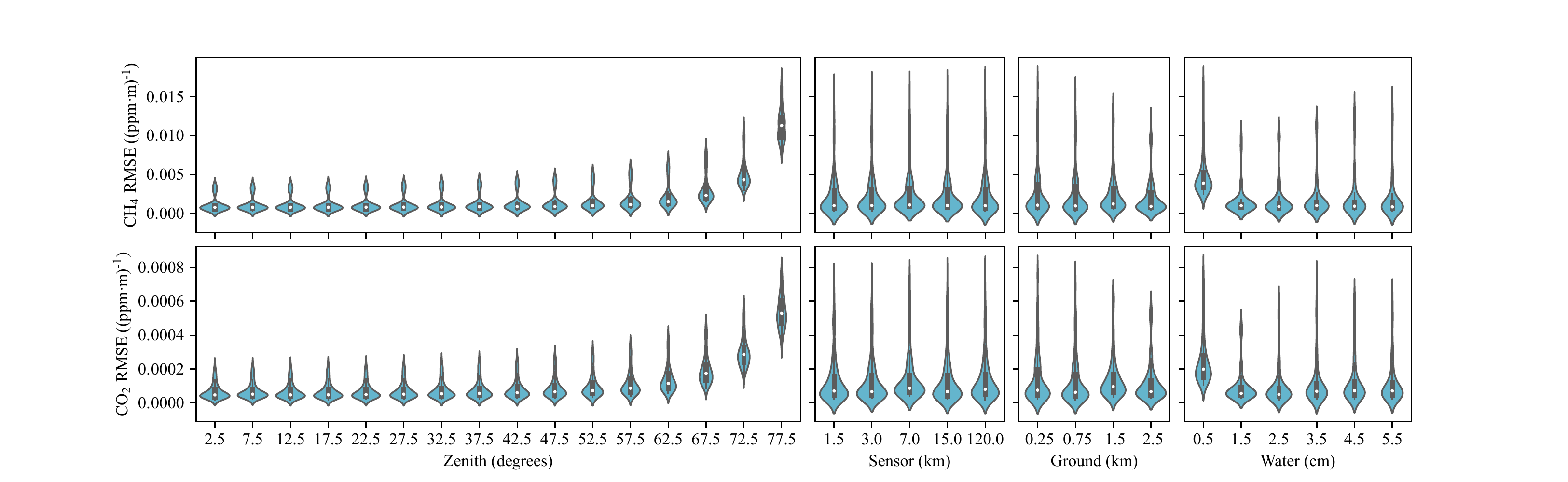}\vspace{-0.45cm}
    \caption{Root mean square error (RMSE) at validation parameter values, which are midpoints between lookup table sample values, are shown as distributions in violin plots. 
    Each violin represents the distribution of RMSE for the population of spectra with the associated validation parameter value. Traditional box plots are inset within each violin, with a white marker for the mean value.
    The error drops to zero at lookup table sampled parameter values (not shown).
    }
    \label{fig:lut_interp_error}
\end{figure*}
Using a set of validation radiative transfer simulations set at half-increments of the sampled grid points of the lookup table, we calculated the error in unit enhancement spectrum generation for a wide set of geometric and atmospheric parameters. 
We expected the half-increment positions to produce interpolated spectra with maximal error, as the linear approximation of a smooth function will deviate the most halfway between samples. For each parameter value, distributions of the root mean squared error between the interpolated and directly simulated unit enhancement spectra are shown in Figure~\ref{fig:lut_interp_error}.
Each distribution includes the population of interpolated spectra with the chosen value of validation parameter on the associated axis and all validation values for other parameters. Outliers at each validation value represent the higher RMSE values from a few samples with other parameter values that are associated with higher error.
The magnitudes of RMSE differed between \chiv{} and \coii{}, which follows from the overall magnitude of \chiv{} unit enhancement spectra being about 10\texttimes{} greater than \coii{} (see Figure~\ref{fig:spectra}).
In the worst case, the root mean squared error was around 0.01~(\ppmm)\textsuperscript{-1} for \chiv{}, and 0.0005~(\ppmm)\textsuperscript{-1} for \coii. 
However, trends of RMSE were similar for \chiv{} and \coii{}.
First, the error was greatest at large zenith angles. 
At these large angles, the path length changes most rapidly for a unit angular shift.
Second, the error was higher at the lowest column water vapor test point value, 0.5 cm. 
In this regime of low column water vapor, the change in radiance from exponential Beer-Lambert absorption is greatest.

Collectively, the lookup table and interpolation approach closely approximates the unit enhancement spectrum from exact parameter simulations for the scenes in the benchmark dataset. As linear interpolation requires access only to the bounding data points, interpolation need not access the full radiance lookup table. 
Our implementation uses this fact to avoid overhead cost when approximating the unit enhancement spectra at arbitrary atmospheric parameters.
Despite the total lookup table being approximately 7~GB, our implementation reads less than 20~MB total from storage when generating the approximate scene-specific unit enhancement spectrum.
Our interpolation implementation uses open-source libraries and file formats for portability and ease of use \citep{Folk2011,Harris2020}.
To measure the change in computational cost between exact simulation of parameters and approximation with the lookup table, we generated unit enhancement spectra for \texttt{ang20191023t151141} using both methods on a Windows~10 virtual workstation with 8 vCPU cores at 2.6~GHz and 64~GB of system memory.
The MODTRAN simulations were performed at 1~cm\textsuperscript{-1} (10 times more coarse than the lookup table), which provides sufficient resolution for simulating AVIRIS-NG data but speeds up radiative transfer simulation.  
These MODTRAN simulations ran for 230.22 seconds.
The corresponding interpolation routine ran for 5.82 seconds.
This 40\texttimes{} speed-up of unit enhancement spectrum generation avoids spectrum generation being the rate-limiting process when performing matched filter processing of a scene, as compared with matched filter processing times reported by \citet{Foote2020} that average to 96 seconds per scene over a different dataset of 300 scenes.
The storage trade-off for this interpolation method, near 7~GB per gas species, is reasonable compared to the storage required for imaging spectrometer scenes from AVIRIS-NG, which typically range from 5--30~GB. 

\begin{figure*}[tbp]
    \centering
    \newlength{\sensaxeslen}
    \setlength{\sensaxeslen}{0.285\linewidth}
    \definecolor{ch}{HTML}{d62728}
    \definecolor{co}{HTML}{1f77b4}
    \footnotesize\scriptsize
    \begin{tikzpicture} 
        \begin{axis}[
            width=\sensaxeslen,
            height=\sensaxeslen,
            xlabel = {(a) Zenith Angle (\degree)},
            ylabel = {\%\,\textDelta\, IME (\% / \textDelta 1 \degree)},
            tick align=outside,
            ytick pos=left,
            xtick pos=lower,
            major tick length=2pt,
            major tick style = {thick, black},
        ]
            \addplot[
                color=ch,
                only marks,
                mark=+
                ] table [
                x=zenith,
                y=normalized_sens_zenith,
                col sep=comma
                ] {figures/data/sensitivity_scatter_ch4.csv};
            \addplot[
                color=co,
                only marks,
                mark=x
                ] table [
                x=zenith,
                y=normalized_sens_zenith,
                col sep=comma
                ] {figures/data/sensitivity_scatter_co2.csv};
        \end{axis}
    \end{tikzpicture}
    \begin{tikzpicture}
        \begin{axis}[
            width=\sensaxeslen,
            height=\sensaxeslen,
            xlabel = {(b) Ground Elevation (km)},
            ylabel = {\%\,\textDelta\, IME (\% / \textDelta 1 km)},
            ytick distance=1,
            tick align=outside,
            ytick pos=left,
            xtick pos=lower,
            major tick length=2pt,
            major tick style = {thick, black},
        ]
            \addplot[
                color=ch,
                only marks,
                mark=+
                ] table [
                x=ground,
                y=normalized_sens_ground,
                col sep=comma
                ] {figures/data/sensitivity_scatter_ch4.csv};
            \addplot[
                color=co,
                only marks,
                mark=x
                ] table [
                x=ground,
                y=normalized_sens_ground,
                col sep=comma
                ] {figures/data/sensitivity_scatter_co2.csv};
        \end{axis}
    \end{tikzpicture}
    \begin{tikzpicture}
        \begin{axis}[
            width=\sensaxeslen,
            height=\sensaxeslen,
            xlabel = {(c) Water Vapor (cm)},
            ylabel = {\%\,\textDelta\, IME (\% / \textDelta 1 cm)},
            tick align=outside,
            ytick pos=left,
            xtick pos=lower,
            major tick length=2pt,
            major tick style = {thick, black},
        ]
            \addplot[
                color=ch,
                only marks,
                mark=+
                ] table [
                x=water,
                y=normalized_sens_water,
                col sep=comma
                ] {figures/data/sensitivity_scatter_ch4.csv};
            \addplot[
                color=co,
                only marks,
                mark=x
                ] table [
                x=water,
                y=normalized_sens_water,
                col sep=comma
                ] {figures/data/sensitivity_scatter_co2.csv};
        \end{axis}
    \end{tikzpicture}
    \begin{tikzpicture}
        \begin{axis}[
            width=\sensaxeslen,
            height=\sensaxeslen,
            xlabel = {(d) Sensor Altitude (km)},
            ylabel = {\%\,\textDelta\, IME (\% / \textDelta 1 km)},
            tick align=outside,
            ytick pos=left,
            xtick pos=lower,
            major tick length=2pt,
            major tick style = {thick, black},
        ]
            \addplot[
                color=ch,
                only marks,
                mark=+
                ] table [
                x=sensor,
                y=normalized_sens_sensor,
                col sep=comma
                ] {figures/data/sensitivity_scatter_ch4.csv};
            \addplot[
                color=co,
                only marks,
                mark=x
                ] table [
                x=sensor,
                y=normalized_sens_sensor,
                col sep=comma
                ] {figures/data/sensitivity_scatter_co2.csv};
            \addlegendentry{\chiv{}}
            \addlegendentry{\coii{}}
        \end{axis}
    \end{tikzpicture}\vspace{-10pt}
    \caption{Sensitivity of \chiv{} and \coii{} IMEs to scene-specific unit enhancement spectrum parameters. Each point represents a single plume from the benchmark dataset.}
    \label{fig:sens}
\end{figure*}
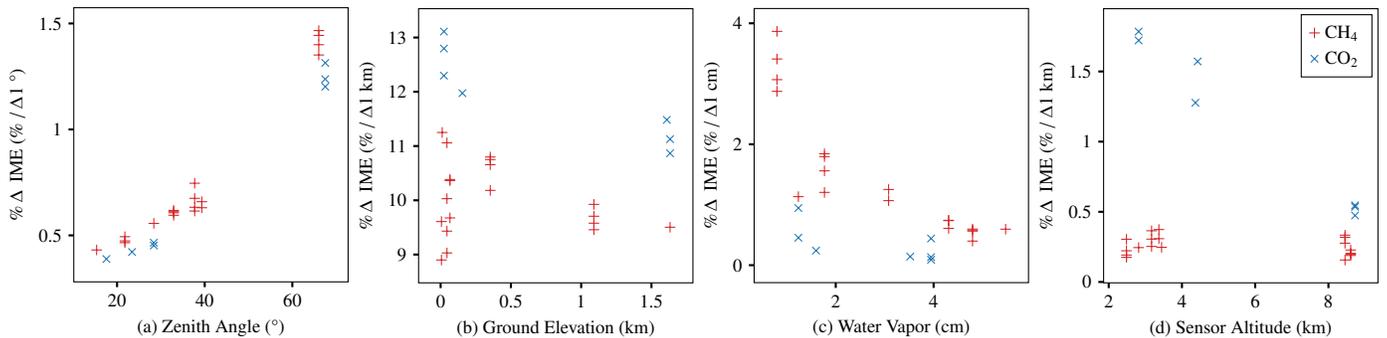

\subsection{Parameter Sensitivity}
\label{sec:results:sens}
Additional gas retrievals and IME calculations using interpolated unit enhancement spectra for parameter value extrema produced the populations of IME changes depicted in Figure~\ref{fig:sens}. 
IME appears to have low sensitivity to solar zenith angle (Figure~\ref{fig:sens}a), with changes of less than 1.5\% IME per 1\textdegree{} change in zenith angle. However, 1\textdegree{} represents a very small increment in solar zenith angle, considering the over 50\textdegree{} range in zenith angles present within the benchmark dataset (Table~\ref{tab:scene_table}). Sensitivity of IME to change in solar zenith angle was especially high at large zenith angles, due to increases in downwelling pathlength approximated by secant of the zenith angle. IME sensitivity to ground elevation was also relatively high, with a 1 km change in ground elevation producing an 8.9 to 13.1\% difference in IME (Figure~\ref{fig:sens}b). Sensitivities of IME to both ground elevation and sensor altitude (Figure~\ref{fig:sens}d) were higher for \coii{}, likely to due to the impact of the high background concentration of \coii{} relative to \chiv{}. IME was less sensitive to changes in water vapor (Figure~\ref{fig:sens}c) and sensor altitude (Figure~\ref{fig:sens}d). IME sensitivity to water vapor was highest for \chiv{} at very low column water vapor values. Water vapor absorption was demonstrated to have an important impact on \chiv{} unit enhancement spectrum shape, and low column water vapor will produce larger changes in absorption based on exponential Beer-Lambert absorption. Sensitivity to sensor altitude was relatively low (particularly for \chiv{}) within the range of altitudes in the benchmark dataset. 

\vfill

\section{Discussion}

Generic enhancement spectra have been extensively used for \chiv{} retrievals, including for studies quantifying emissions flux \citep{Thompson2015,Frankenberg2016,Thompson2016,Krautwurst2017,Duren2019,Thorpe2020}. Due to the balance between relatively weak background \chiv{} absorption and accounting for two-directional transmittance, errors due to use of a generic enhancement spectrum for \chiv{} were not uniformly positive or negative and vary by scene and by background surface reflectance. Although average \chiv{} IME error due to the generic unit enhancement spectrum was low, the error for individual plumes may rival the uncertainties introduced by wind speed and mass integration for some plumes. \citet{Duren2019} calculated a 29\% mean uncertainty due to mass integration and a 22\% mean uncertainty due to wind speed across hundreds of \chiv{} plumes in California. Use of scene-specific \chiv{} enhancement spectra should help reduce overall error in flux estimates, although the assumption of representing enhancements as a uniform layer does need to be validated. The generic unit enhancement spectrum approach will be least accurate for scenes with longer pathlengths, so geometric and atmospheric effects should be accounted for when working with scenes with high solar zenith angles and large ground-to-sensor distances (including satellite data). 

Not accounting for relatively high background levels of \coii{} causes the generic \coii{} unit enhancement spectrum to have a much higher magnitude than spectra simulated using radiative transfer modeling. As a consequence, \coii{} IME and flux will be greatly underestimated using a generic spectrum. Studies to date that have utilized generic enhancement spectra have only considered \chiv{}, but given our findings, generic enhancement spectra generated using the \citet{Thompson2016} approach should not be used for \coii{} retrievals. 

Column water vapor had an important impact on the shape of scene-specific unit enhancement spectra. Both \chiv{} and \coii{} overlap with water vapor absorption in the shortwave infrared. Inclusion of water vapor absorption changed the spectral shape of enhancement spectra in comparison to the generic spectra, resulting in reduced magnitude at long wavelengths on the \chiv{} continuum and at short wavelengths on the \coii{} continuum. Since imaging spectroscopy can provide estimates of column water vapor in addition to \chiv{} and \coii{} retrievals \citep{Thompson2015a}, scene-specific or even intrascene corrections for water vapor absorption should be readily achievable. The combined effects of water vapor, pathlength, and background \chiv{} absorption resulted in changes in scene-specific unit enhancement spectrum shape that could not be easily accounted for without radiative transfer modeling.  

This analysis did not examine impacts of varying aerosol optical depth, atmospheric model, and view zenith angle on unit enhancement spectrum modeling. Initial empirical analysis has shown that these parameters produced much lower impacts on \chiv{} and \coii{} enhancement spectra than the solar zenith angle, ground elevation, sensor altitude, and water vapor parameters used in this study. Aerosol optical depth is low in the shortwave infrared, and view zenith angle for AVIRIS-NG is constrained to less than 18\textdegree{} by the instrument’s field of view. However, a mismatch in atmospheric model between the tropical profile used for the India scenes and the mid-latitude summer profile used for the lookup table did demonstrate a small difference in IME. These atmospheric models differ in their temperature and density profiles, and future work should examine the importance of these profiles, along with accounting for aerosol scattering and view zenith angle, in further reducing error in IME calculation and flux estimation. 

Our approximation technique using interpolation on a precomputed lookup table of radiative transfer simulations sacrificed only a small degree of confidence, as long as simulations with the same atmospheric model were compared. Although specific to the lookup table approach, the sampling strategy we used leaves some parameter values with higher interpolation error, particularly at large solar zenith angle and low column water vapor. The error could be further reduced by including additional simulations in the lookup table representing large solar zenith angles and low water vapor. The lookup table was produced at suitable resolution for application to current imaging spectrometers and potential future instruments with increased spectral resolution, but may need to be updated for changing background concentrations of \coii{} and \chiv{} in the future. Our sensitivity analysis demonstrated relatively low sensitivity of plume IME to changes in column water vapor and sensor altitude. Depending on expected ranges in these parameters, future methods for estimating scene-specific unit enhancement spectra may be able to exclude these two parameters with a relatively small impact on accuracy. 

It is important to note that none of the benchmark imaging spectrometer scenes contain cloud cover. Robust cloud screening is assumed for application of matched filter retrievals of \chiv{} and \coii{}, and impacts of partial cloud cover on retrievals are unknown. 

Unit enhancement spectra using scene-specific geometric and atmospheric parameters should provide more accurate IME values and flux estimations. However, since there are no `true' concentration enhancements or fluxes for the \chiv{} plumes in the benchmark dataset, simulated scenes, such as those used in \citet{Foote2020}, may be needed to demonstrate accuracy gains over generic unit enhancement spectra. Since simulated scenes are constructed using radiative transfer modeling, it follows that unit enhancement spectra derived from radiative transfer would prove to be more accurate when applied to simulated scenes. Controlled \chiv{} release experiments may provide an alternative method for assessing the relative accuracy of scene-specific and generic unit enhancement spectra for \chiv{} flux estimation \citep{Thorpe2016}.

\section{Conclusion}
The use of generic unit enhancement spectra for concentration-pathlength retrieval can lead to errors in IME calculation and flux estimation. Radiative transfer simulation of at-sensor radiance can account for full downwelling and upwelling paths, including solar zenith angle and ground-to-sensor distance. Our scene-specific unit enhancement spectrum approach also accounts for the background concentrations of \chiv{} and \coii{} in the atmosphere, along with water vapor. Modeling these parameters resulted in important differences in magnitude and spectral shape of scene-specific unit enhancement spectra. Examination of ten benchmark scenes containing \chiv{} and \coii{} plumes found IME differences ranging from $-$22.0\% to 28.7\% for \chiv{}, and from $-$76.1\% to $-$48.1\% for \coii{}. These percentages correspond to differences in point source flux assuming that wind speed and plume length are held constant for each plume and only the unit enhancement spectrum is varied.  

Errors in \chiv{} IME caused by geometric effects and the background concentrations of trace gases are easily addressable with minimal computational cost by using interpolated, scene-specific unit enhancement spectra. The average $-$2.3\% error in \chiv{} IME produced by the application of generic unit enhancement spectra to the benchmark dataset indicates that when averaged across a large range of geometric and atmospheric parameters, errors in \chiv{} IME and fluxes estimated in previous studies may be relatively low. However, more limited ranges in parameters or more extreme parameters within individual campaigns may produce larger average errors than those shown here. At the level of a single plume, error in \chiv{} IME and flux from using a generic unit enhancement spectrum may exceed $\pm$20\% and be dependent on background surface reflectance.  These effects from the modified unit enhancement spectrum are expected from all matched filter algorithms, beyond the single matched filter variant we investigated. 

Based on the large differences between generic and scene-specific unit enhancement spectra for \coii{}, and resulting large differences in calculated IME, the approach currently used for generating generic unit enhancement spectra for matched filter methods does not appear to be suitable for \coii{} flux estimation. However, our results do not preclude use of radiative transfer modeling to produce a new generic spectrum that is more representative of \coii{} or \chiv{} absorption. While sacrificing the ability to correct for scene-specific variation in parameters, such an approach could still greatly improve \coii{} flux estimation using quantitative matched filters. 

A large number of satellite imaging spectrometer missions capable of mapping \chiv{} and \coii{} point source emissions have recently began operation or are in development, including PRISMA \citep{Loizzo2019}, EnMAP \citep{Guanter2015}, EMIT \citep{Bradley2020}, SBG \citep{2019AGUFMGC54A..04P}, and CHIME \citep{Nieke2018}. Future work should evaluate the utility of both scene-specific and intra-scene correction of unit enhancement spectra with matched filter approaches, as well as compare results from matched filter approaches to results from DOAS \citep{Ayasse2019,Cusworth2019}.

\appendix
\section*{Acknowledgements}
We appreciate the assistance of the AVIRIS-NG team, including Winston Olson-Duvall and John Chapman, in constructing the benchmark dataset. AVIRIS-NG flights were funded by the California Air Resources Board, the California Energy Commission, and NASA’s Earth Science Division. 
Open-source software packages were valuable in producing this work, including \texttt{matplotlib} \citep{Hunter:2007}, \texttt{pandas} \citep{mckinney-proc-scipy-2010,jeff_reback_2021_4681666} and \texttt{seaborn} \citep{Waskom2021}.
We appreciate the insightful comments of three anonymous reviewers, which allowed us to improve the manuscript.
\section*{Funding}
Funding for this research was provided by NASA Carbon Monitoring System grant 80NSSC20K0244.

\def\bibfont{\scriptsize}
\bibliographystyle{elsarticle-harv}
\bibliography{target_spectrum.bib}

\end{document}